# The isotope effect: Prediction, discussion, and discovery

Helge Kragh[*]

Centre for Science Studies, Department of Physics and Astronomy, Aarhus University, 8000 Aarhus, Denmark.

ABSTRACT

The precise position of a spectral line emitted by an atomic system depends on the mass of the atomic nucleus and is therefore different for isotopes belonging to the same element. The possible presence of an isotope effect followed from Bohr's atomic theory of 1913, but it took several years before it was confirmed experimentally. Its early history involves the childhood not only of the quantum atom, but also of the concept of isotopy. Bohr's prediction of the isotope effect was apparently at odds with early attempts to distinguish between isotopes by means of their optical spectra. However, in 1920 the effect was discovered in HCl molecules, which gave rise to a fruitful development in molecular spectroscopy. The first detection of an atomic isotope effect was no less important, as it was by this means that the heavy hydrogen isotope deuterium was discovered in 1932. The early development of isotope spectroscopy illustrates the complex relationship between theory and experiment, and is also instructive with regard to the concepts of prediction and discovery.

## 1. Introduction

The wavelength of a spectral line arising from an excited atom or molecule depends slightly on the isotopic composition, hence on the mass, of the atomic system. The phenomenon is often called the "isotope effect," although the name is also used in other meanings. Ever since 1920, when this effect was first discovered in the band spectra of simple molecules, it has played an important scientific role and also been applied to a variety of problems of an applied nature. Today the isotope effect spans several sciences, not only physics and chemistry but also astronomy, geology, biology, and the environmental sciences.[1]

[*] E-mail: helge.kragh@ivs.au.dk.

The aim of this paper is to discuss the early history of the isotope effect, mainly in the period from about 1913 to 1921. It also includes a brief account of the first detection of an atomic isotope shift, which occurred in connection with the important discovery of deuterium in 1932. While the discoveries of isotope effects in band spectra and atomic spectra (in 1920 and 1932, respectively) are well known, it is not generally known that the effect was predicted by Niels Bohr shortly after the publication of his atomic theory in the summer of 1913. Nor was this well known at the time. The paper pays particular attention to Bohr's ideas about the subject and the role that the isotope effect played in the early phase of the Bohr quantum atom.

**2. Isotopes and atomic structure**

None of the early theories of atomic structure that were developed in the first part of the twentieth century foresaw the possibility of species of the same chemical element with different atomic weights. Yet the idea was anticipated by the chemist William Crookes as early as 1886 in a far-ranging address to the British Association for the Advancement of Science. At this occasion he suggested that "when we say that the atomic weight of, for instance, calcium is 40, we really express the fact that, while the majority of the calcium atoms have an actual weight of 40, there are not a few which are represented by 39 or 41, a less number by 38 and 42, and so on" (Crookes, 1886, p. 569). He also speculated that the still hypothetical helium, supposed to be of atomic weight 0.5, might be the prime matter out of which all the elements had been formed in a cosmic evolution process. Crookes's spirited idea led him to interesting speculations about "meta-elements," but it turned out not to be viable and was only resuscitated some 25 years later.

---

[1] See Wolfsberg et al. (2010) for a comprehensive and historically oriented survey. There exists no account of the isotope effect in the history of science literature.

It was primarily the perplexing study of radioactive decay series that first indicated the possibility of isotopy, atoms with a few different physical properties belonging to the same element.[2] Another source was the experiments with positive rays or "canal rays" that J. J. Thomson had conducted in Cambridge since 1906 and eventually would be developed into the powerful technique of mass spectroscopy (Falconer, 1988; Budziekiewicz and Grigsby, 2006).

Some of the substances found in the radioactive series and identified by their radioactive properties turned out to have a strong chemical resemblance to other elements; in fact, they were inseparable from them and yet they were not identical to them. In desperation, some scientists grouped several radio-elements (such as radium emanation, actinium emanation, and thorium emanation) into the same place in the periodic system, while others suggested to extend the periodic system to accommodate the new radio-elements. By 1911 Rutherford's former collaborator Frederick Soddy, then at the University of Glasgow, concluded that radium, mesothorium 1, and thorium X, although of different atomic weights and radioactive properties, were not merely chemically similar, but chemically identical. The strange phenomenon might be a peculiarity of the radioactive elements in the upper part of the periodic system, but Soddy thought this was not the case. According to him, it was likely to be generally valid. When he coined the word "isotope" in late 1913, he related it to Rutherford's nuclear atom:

> The same algebraic sum of the positive and negative charges in the nucleus, when the arithmetical sum is different, gives what I call "isotopes" or "isotopic elements", because they occupy the same place in the periodic table. They are chemically identical, and save only as regards the relatively few physical properties which depend upon atomic mass directly, physically identical also. Unit changes of this nuclear charge, so reckoned algebraically, give the successive places in the periodic table. For any one "place" or any one nuclear charge, more than one number of electrons in the outer-ring system may exist, and in such a case the element exhibits variable valency.[3]

---

[2] For the he history of isotopy in the period 1910-1915, see Brock (1985, pp. 196-216), and Bruzzaniti and Robotti (1989). Soddy's series of progress reports on radioactivity to the Chemical Society 1904-1920 is a valuable primary source of information (Trenn, 1975).

The Polish physical chemist Kasimir Fajans, who had spent the year 1910-1911 working with Rutherford in Manchester, had in an earlier paper (Fajans, 1913) suggested essentially the same hypothesis. He called a group of chemically identical elements a "pleiade," but, while "isotope" caught on, "pleiade" did not. Apart from Fajans himself, very few scientists used the term, which was soon forgotten.[4] Contrary to Soddy, at the time Fajans did not accept Rutherford's nuclear atom, but argued that alpha particles were expelled from the outer layer of the atom. He only came to accept the nuclear model after it had been extended to the Bohr-Rutherford theory of atomic structure. As he wrote to Rutherford in a letter of 13 December 1913: "I have followed Bohr's papers with extraordinary interest, and now I no longer doubt the complete correctness of your atomic theory. The reservations I expressed in my last letter have been entirely removed by Bohr's work" (quoted in Jensen, 2000, p. 34).

---

[3] Soddy (1913, p. 400), issue of 4 December. Reprinted together with other historical papers on radiochemistry and isotopes in Romer (1970, pp. 251-252). The "positive and negative charges in the nucleus" is a reference to protons and electrons.

[4] The term was used by Sommerfeld in his influential *Atombau und Spektrallinien*, but in Fajans's sense of a group of isotopes belonging to the same element

Radioactive decay was not the only phenomenon that pointed towards isotopy. So did the positive rays, or what in Germany were known as *Kanalstrahlen*, that were investigated by J. J. Thomson, Wilhelm Wien, Johannes Stark, and others. Francis Aston, who served as Thomson's assistant in parts of his research programme, analyzed positive rays of neon, known to have the atomic weight 20.2. Surprisingly, Aston's experiments revealed not only rays corresponding to atomic weight 20 but also weaker rays corresponding to atomic weight 22. In lack of a proper explanation, he suggested to have found what he called "meta-neon," possibly a new inert gas (Aston, 1913; see also Brock, 1985, pp. 205-215.). After a brief period of confusion, Aston and Soddy realized that what Aston had discovered was a heavy isotope of neon. Thomson at first thought that the recorded species of atomic weight 22 might be the neon hydride compound $NeH_2$ and only reluctantly agreed that neon was probably a mixture of chemically inseparable species with different atomic weights. Having his own ideas of atomic constitution, he did not agree with the interpretation of isotopy in terms of the nuclear model of the atom. As late as May 1921 Thomson

---

(Sommerfeld, 1922, p. 103). Note that the terms "pleiade" and "isotope" were not equivalent. Fajans did not suggest a name for a particular species of a chemical element.

suggested that some of Aston's results might be explained by the formation of hydrides rather than separation into isotopes (Thomson et al., 1921).

Although the name "isotope" may have come as a surprise to Bohr, the concept did not. In an interview shortly before his death in 1962, he even claimed that "it was really me who got the idea of isotopes."[5] Whatever the credibility of this recollection, Bohr had for some time suspected that the chemical elements might exist in versions with different atomic weights and nuclear structures that might account for the confusing radioactive properties of the heavy elements. Thus, it was known that some radioactive substances, apparently belonging to the same element, emitted beta rays with different velocities. According to Bohr, these substances had the same electron systems and only differed in their atomic weights, meaning their nuclei. The phenomenon provided strong support of "the hypothesis that the β–rays have their direct origin in the nucleus," as he wrote to Rutherford in June 1913.[6] A similar formulation appeared in the second part of his great trilogy published in the October 1913 issue of *Philosophical Magazine*. Referring to substances that "are different only in radio-active properties and atomic weight but identical in all other physical and chemical respects," he concluded: "The charge on the nucleus, as well as the configuration of the surrounding electrons, [is] identical in some of the elements, the only difference being the mass and the internal constitution of the nucleus" (Bohr, 1913a, p. 501).

Looking back on the development many years later, in his Rutherford Memorial Lecture of 1958 he recalled (Bohr, 1961, p. 1085):

---

[5] Interview with Bohr of 1 November 1962, by T. S. Kuhn, L. Rosenfeld, Aa. Petersen, and E. Rüdinger. Transcript by Niels Bohr Library & Archives, American Institute of Physics, http://www.aip.org/history/ohilist/4517_1.html.
[6] Bohr to Rutherford, 10 June 1913, in Bohr (1981, p. 586).

> When I learned that the number of stable and decaying elements already identified exceeded the available places in the famous table of Mendeleev, it struck me that such chemically inseparable substances, to the existence of which Soddy had early called attention and which later by him were termed "isotopes," possessed the same nuclear charge and differed only in the mass and intrinsic structure of the nucleus.

Young Bohr wanted to write a paper about his ideas of what later became isotopy and the atomic number, but was dissuaded by Rutherford: "When I turned to Rutherford to learn his reaction to such ideas, he expressed, as always, alert interest in any promising simplicity but warned with characteristic caution against overstressing the bearing of the atomic model and extrapolating from comparatively meager experimental evidence" (Bohr, 1961, p. 1085).

### 3. Bohr's predictions

Aston's announcement of "meta-neon" took place at the September 1913 meeting in Birmingham of the British Association for the Advancement of Science, the same meeting where Bohr's atomic theory was first introduced and discussed by leading physicists. In Birmingham, Bohr listened not only to Aston's presentation but also to Thomson's talk with the enigmatic title "$X_3$ and the Evolution of Helium" in which Thomson discussed his discovery of what he argued was triatomic hydrogen ($H_3$ and $H_3^+$) in experiments with positive rays. Although unable to obtain spectroscopic evidence for the unusual $H_3$ molecule, he was convinced that it was stable and nearly as chemically inert as the noble gases (Thomson, 1913; Kragh, 2011).

Bohr found the $H_3$ hypothesis interesting, but in the discussion following Thomson's talk he suggested as an alternative the bold hypothesis that $X_3$ might possibly be a heavy isotope of hydrogen of atomic weight 3. That is, he effectively predicted what was later called tritium or hydrogen-3. Nearly fifty years later Bohr recalled the incident as follows: "I just took up the question of whether in hydrogen one could have what you now call tritium. And then I saw that it was a way to show this by its diffusion in palladium. Hydrogen and tritium will behave similarly but the masses are so different that they will get separated out."[7]

As we know from a letter from the Hungarian chemist George de Hevesy to Rutherford, Bohr's suggestion of "an H atom with one central charge, but having a three times so heavy nucleus than Hydrogen" was not taken seriously.[8] Yet, after having returned to Copenhagen he continued to think about it, realizing that in principle the question might be resolved by spectroscopic precision experiments. According to Bohr's theory of one-electron atoms (Bohr, 1913a), the wave number $1/\lambda = \nu/c$ corresponding to a quantum transition from state $n_1$ to $n_2$ could be written as

$$\frac{1}{\lambda} = Z^2 R_\mathrm{H} \left(\frac{1}{n_2^2} - \frac{1}{n_1^2}\right),$$

where Rydberg's constant for hydrogen is given by

$$R_\mathrm{H} = \frac{2\pi^2 m e^4}{h^3 c}$$

---

[7] Interview with Bohr of 1 November 1962 (Niels Bohr Library & Archives, American Institute of Physics, http://www.aip.org/history/ohilist/4517_1.html). Bohr also recalled: "When Rutherford came to Cambridge [in 1919], Thomson wouldn't even believe in the isotopes; he didn't believe in Aston's experiments." It had been known since about 1870 that palladium is able to absorb large volumes of hydrogen and that the process depends critically on the purity (molecular weight) of the hydrogen gas.

[8] Hevesy to Rutherford, 14 October 1913, quoted in Eve (1939, p. 224). Hevesy thought that Bohr's hypothesis "is not very probable, but still a very interesting suggestion, which should not be quickly dismissed." A specialist in radiochemistry, Hevesy knew Bohr from his stay with Rutherford in Manchester, where he was an important source for Bohr's ideas about radioactivity.

The symbol $m$ denotes the mass of the electron, $Z$ is the nuclear charge, and the other symbols have their usual meanings. In the first part of his 1913 trilogy Bohr had brilliantly shown how the expression could be rearranged so at to account for the so-called Pickering-Fowler lines, which according to Bohr had their origin in He$^+$ ($Z$ = 2) and not in H ($Z$ = 1) as commonly assumed. However, although the British spectroscopist Alfred Fowler reluctantly acknowledged Bohr's explanation – or "monster-adjustment," as Imre Lakatos (1970) later called it – he was not yet satisfied. In the fall of 1913 he pointed out that although Bohr's theoretical values for $\lambda_\mathrm{H}$ and $\lambda_{\mathrm{He}^+}$ were very close to the observed wavelengths, they did not quite agree with them (Fowler, 1913). Bohr's response to the new challenge was to modify his theory by taking into account the finite mass of the nucleus, which he did by replacing the electron mass $m$ with the reduced mass $\mu$ given by

$$\mu = \frac{mM}{m+M} = \frac{m}{1+m/M} \quad \text{or} \quad \frac{1}{\mu} = \frac{1}{m} + \frac{1}{M},$$

where $M$ is the mass of the nucleus. In this way the Rydberg constant would depend slightly on the mass of the element. As Bohr (1913b; 1914) demonstrated, with the replacement the discrepancies mentioned by Fowler disappeared. "The agreement with the experimental value [of $R_H$] is within the uncertainty due to experimental errors in $h$, $e$, and $m$," he pointed out (Bohr 1913b, p. 232). While $R_H$ = 109 675 cm$^{-1}$ (Bohr's value), the value for the heavier elements would be close to

$$R_\infty = R_H \left(1 + \frac{m}{M_H}\right) = 109\ 735\ \text{cm}^{-1}$$

For the sake of comparison, to the same accuracy the modern values are $R_\infty$ = 109 737 cm$^{-1}$ and $R_H$ = 109 678 cm$^{-1}$.

Bohr realized that the same kind of reasoning could be applied to the spectral lines of isotopes, yielding an isotope shift for the hypothetical hydrogen-3 (or tritium, T) of

$$\Delta \lambda = \left(1 - \frac{R_H}{R_T}\right) \lambda_H = \frac{2m}{3m + 3M} \lambda_H \cong 3.6 \times 10^{-4} \lambda_H$$

In the case of hydrogen-2 (which Bohr did not consider), the result becomes $\Delta \lambda \cong 2.7 \times 10^{-4} \lambda_H$ cm$^{-1}$. Together with a colleague in Copenhagen, the spectroscopist Hans Marius Hansen, Bohr examined the question from both a theoretical and experimental point of view. In an unpublished note on "Spectrum of Hydrogen Isotope" from either late 1913 or early 1914 (Bohr, 1981, p. 416), he wrote about the two possibilities of Thomson's positive ray particle of mass 3:

> A theory of spectra proposed by one of us offer a possibility to discriminate between the above eventualities, since according to this theory the spectrum of an element of atomic weight 3 and, as hydrogen, containing one electron should show a spectrum the lines of which are closely but not exactly coinciding with the lines of the hydrogen spectrum; the difference in wave length being of an order open for detection.

The experiments conducted in Copenhagen were unsuccessful, which was presumably the reason why Bohr chose not to publish his considerations. Incidentally, also later attempts to detect tritium by means of the isotope effect failed. Following the discovery of deuterium in 1932, Gilbert Lewis and Frank Spedding at the University of California looked for the superheavy isotope in the solar spectrum and also in heavy water, but without finding it. All they could conclude was that "ordinary hydrogen can not contain more than one part in six million of $H^3$" (Lewis and Spedding, 1933, p. 964). When tritium was eventually discovered (Alvarez and Cornog, 1939), it was as a β-radioactive substance produced artificially by bombarding deuterium with deuterons accelerated in a cyclotron:

$$^2H + {^2H} \rightarrow {^1H} + {^3H} \quad \text{followed by} \quad ^3H \rightarrow {^3He} + \beta + \bar{\nu}.$$

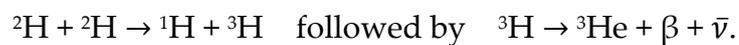

It should be pointed out that it is somewhat debatable when tritium was discovered and who made the discovery (Eidinoff, 1948). The same deuteron-deuteron reaction had been reported by Rutherford and his two collaborators Mark Oliphant and Paul Harteck in 1934. In his last scientific paper Rutherford (1937) carefully evaluated the evidence for and against tritium, concluding that although the nuclei of the isotope (tritons) had been detected in nuclear reactions, tritium had not been obtained in such quantities that its properties could be studied by ordinary physical and chemical methods.

In a paper of 4 March 1915, the British physicist Frederick Lindemann (1915) suggested that if the atomic volumes of isotopes were the same, some of their physical properties should differ because of the different configurations of the nuclei. He predicted that in some cases differences in the melting points, elastic constants, and vapour pressures should be measurable. Bohr drafted a letter of response to *Nature*, but for unknown reasons decided not to send it. In this draft he mentioned about the atomic isotope effect in the visible range that "it would be far too small to be detected by the present means." But he also noted that "the question is somewhat different" for molecules formed by isotopes in the solid state. For in this case "the frequencies with which the atoms vibrate in the molecules would not be the same but would be inversely proportional to the square root of the atomic weight" (Bohr, 1981, p. 416).

Bohr first mentioned the isotope effect in public at the September 1915 meeting of the British Association in Manchester. According to the report in *Nature*: "Dr. N. Bohr pointed out that … in the case of spectral vibrations, there occurs a small term depending on the mass of the central nucleus, and accordingly we ought to look out for a small but perceptible difference between the spectra of two isotopes."[9] In notes written shortly after the meeting, Bohr was more specific:

---

[9] Report of the British Association meeting in Manchester, in *Nature*, 96 (1915), p. 240. Bohr did not present an address in Manchester, and he did not directly refer to the isotope effect in any of his publications at the time. Consequently, few people knew of his ideas.

> We shall expect a correction of the wave-lengths represented by the factor $1 + m/M$, where $m$ and $M$ are the masses of the electron and the nucleus respectively. For elements of high atomic weight this correction is very small, e.g. the difference in the correction for isotopic lead of atomic weight 206 and 208.4 is only $3 \times 10^{-8}$. This is consistent with the recent experiments of Merton …, which show that a possible difference in the wave-lengths for Uranium and Thorium lead at any rate is less than $10^{-6}$. On the other hand, the difference to be expected for the two isotopic neons of atomic weight 20 and 22 is $2.5 \times 10^{-6}$ and might perhaps be detectable.[10]

He further predicted an isotopic shift in the spectra of diatomic molecules consisting of atoms of different masses $M_1$ and $M_2$, such as HCl and HBr. In this case the frequency of vibration in the infrared bands would be proportional to the quantity

$$\nu_{vib} = \nu_0 \sqrt{\frac{M_1 + M_2}{M_1 M_2}}$$

"We shall therefore expect appreciable differences in the frequencies of the ultrared absorption bands," Bohr wrote. Thus, in the case of HCl a small isotopic shift should occur, revealing the different masses of the two chlorine isotopes Cl-35 and Cl-37. However, this was not known at the time. It would take five more years until the isotopic composition of chlorine became known (Section 5).

---

[10] "Note on the Properties of Isotopes and the Theory of the Nucleus Atom." Draft of October 1915, in Bohr (1981, pp. 417-425), quotation on p. 419.

## 4. Attempts to detect the isotope effect

Bohr's prediction of an isotope effect was not followed up immediately, but in 1917 Lester Aronberg at the University of Chicago investigated a possible isotope shift in the case of the lead isotopes Pb-207 and Pb-206, of which the latter at the time was known as radio-lead or Radium G (Harkins and Aronberg, 1917; Aronberg, 1918; see also Aston, 1923, pp. 123-125). Several physicists and chemists had previously looked for differences in the spectra of isotopes, and they had all concluded that the spectra were identical within the limits of accuracy. Aronberg's measurements were motivated by theoretical speculations of the physical chemist William Harkins, also at the University of Chicago, who suspected that the mass of the nucleus might have an effect on what he called the period of vibration of extranuclear electrons. (Harkins also speculated that the H-3 isotope, or what he called "eka-hydrogen," might be a constituent of atomic nuclei.) Examining the line $\lambda$ = 4058 Å from ordinary lead, the two Chicago scientists reported a shift in wavelength for Pb-206 of

$$\Delta \lambda = \lambda_{206} - \lambda_{207} = 0.0044 \text{ Å}$$

As they pointed out, Bohr's formula also resulted in a positive shift, but for $Z$ = 82 it would only be $5 \times 10^{-5}$ Å, about 100 times smaller than the value found experimentally.

The Aronberg-Harkins result was confirmed by the Oxford spectroscopist Thomas Ralph Merton (1920) using a different experimental technique. According to him, the shift in wavelength between the two species of lead was $\Delta\lambda = (0.0050 \pm 0.0007) \times 10^{-5}$ Å. He also found a difference in the wavelengths for ordinary thallium (a mixture of Tl-203 and Tl-205) and for thallium obtained from pitchblende (Tl-206) of the same magnitude. Whereas Merton did not relate his result to Bohr's formula or otherwise tried to explain it, in 1920 Harkins and Aronberg drew attention to the discrepancy between theory and experiment. They apparently thought to have "discovered an effect upon the vibration of an electron" and speculated that this effect was "related in some way to the electromagnetic field in the atom, which is related in some unknown way to the large scale gravitational field" (Harkins and Aronberg, 1920, p. 1333).

Harkins suggested that his work with Aronberg might somehow throw light on the smallness of the gravitational force as given by the ratio $4.1 \times 10^{-40}$ between this force and the electrostatic force. He was not the only one at the time to consider the ratio a fascinating theoretical challenge, perhaps an indication of a bridge between Einstein's new gravitation theory and electromagnetism. So did, if from a very different perspective, Hermann Weyl and Arthur Eddington.[11]

Although the magnitude of the isotope shift in lead observed by Aronberg, Harkins, and Merton disagreed violently with the one derived from Bohr's theory, the three authors realized that no firm conclusion could be drawn from the result. After all, the Bohr formula presupposed a one-electron atomic system, and there was no reason why it should be valid also for the much more complex lead atom. The problem of applying Bohr's theory of isotopes to many-electron atoms was discussed in a note by Paul Ehrenfest (1922) to which Bohr added a postscript. Bohr thought that "without a closer investigation" – one of his favourite expressions – his atomic theory could say nothing definite about the isotope effect in heavy atoms. According to his new ideas of atomic structure, some of the electrons would penetrate in their elliptic orbits into the inner parts of the atom which would make calculations impossibly complicated. There might be an isotope effect, but it was hardly possible to say whether it was "sufficiently large to account for the discrepancies observed by Merton, in the wave-lengths of certain lines in the spectra of lead isotopes, which although very small are yet much larger than those to be expected from the simple formula" (Bohr, in Ehrenfest, 1922).

---

[11] What is sometimes called "Weyl's number," $e^2/GmM \cong 10^{40}$, can be expressed as the ratio of the electrical to the gravitational force between an electron and a proton. It it usually traced back to Weyl (1919), but can be found earlier. For example, Richardson (1914, p. 590) called attention to "the smallness of gravitational attraction compared with the forces between the electrons composing the attracting matter," for which he gave the number $4 \times 10^{-40}$. Harkins and Aronberg (1920) referred to Richardson's book.

The question might seem to be more manageable in the simple case of lithium, which was known to exist in the two isotopic forms Li-6 and Li-7 found by Aston. Together with other issues of isotopy, the spectroscopic isotope shift was discussed in a meeting of 2 May 1921 in the Royal Society with contributions from Thomson, Merton, Aston, Soddy, and Lindemann (Thomson et al., 1921). From Bohr's formula applied to $Li^{2+}$ Merton found for the doublet at $\lambda = 6708$ Å that it should be isotopically displaced by 0.087 Å. Such a difference in wavelength "should easily be visible," he said. However, he was unable to find any experimental evidence in the lithium spectrum for components corresponding to the expected isotope effect, and for this reason its existence was still to be considered hypothetical.

## 5. A most beautiful confirmation

In the years after 1918 molecular spectroscopy evolved into a major research field, cultivated in particular by German and American physicists and physical chemists.[12] With the authoritative status that Bohr's atomic theory had achieved, it became important to base a theory of the band spectra fully on this theory, meaning that all lines in whatever range of the spectrum had to represent a transition between quantized energy states. Work within this framework was vigorously pursued by, among others, Fritz Reiche and Adolf Kratzer in Germany, Edwin Kemble in the United States, and Torsten Heurlinger in Sweden.

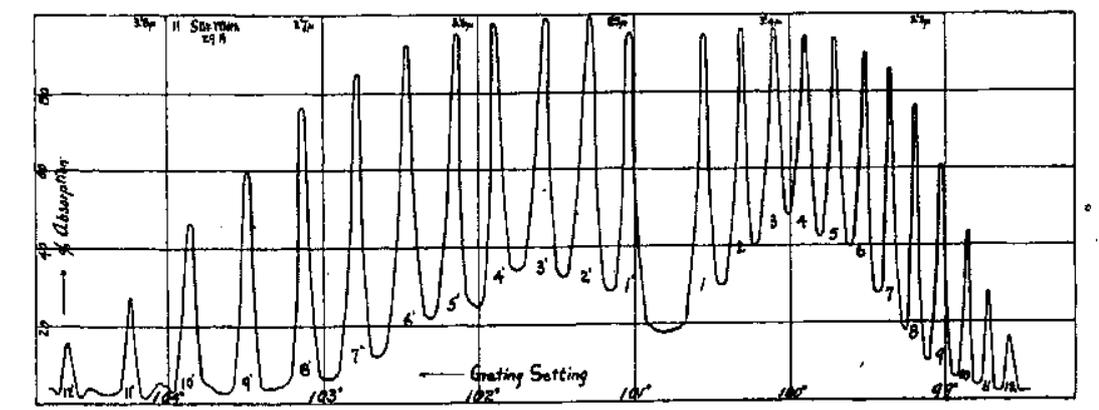

Figure 1. Imes's absorption spectrum of HCl at $\lambda = 3.46\mu$, showing the gap in the pattern of lines that forced physicists to consider half-integral quantum numbers.

A full integration of Bohr's theory into the theory of band spectra occurred in the years 1920-1922, when the frequency condition $\Delta E = h\nu$ was applied to combined electronic, vibrational, and rotational transitions. Through this work it became understood that the energy of a molecule can be divided into three spectral ranges: In the far infrared rotation spectra dominate, while in the near infrared one observes vibration-rotation spectra; the visible and ultraviolet parts of the spectrum are characterized by lines originating in electronic transitions.

---

[12] There is no comprehensive work on the early development of molecular spectroscopy, but valuable information can be found in Fujisaki, 1983, Assmus, 1992a, and Assmus, 1992b. For a contemporary discussion, see for example Sommerfeld (1922, pp. 505-551).

In a paper of 1920 Reiche argued that the frequencies of vibration-rotation spectra were given by the expression

$$\nu = \nu_{vib} \pm \left(m + \frac{1}{2}\right)\frac{h}{4\pi^2 J},$$

where $m$ = 0, 1, 2, … and $J$ is the inertial moment of the molecule (Reiche, 1920). However, it then followed that the lines in a band were evenly spaced with $\Delta\nu = h/4\pi^2 J$, and this implication was contradicted by precision experiments on the absorption in HCl and HBr made by Elmer Imes at the University of Michigan (Imes, 1919, dated 30 April 1918). Imes's measurements showed very clearly a gap in the centre of the pattern of lines that could not easily be explained theoretically (Figure 1). The problem caused Reiche to change the expression for the energy of a rotator from

$$E_{rot} = m^2 \frac{h^2}{8\pi^2 J}$$

to

$$E_{rot} = \left(m + \frac{1}{2}\right)^2 \frac{h^2}{8\pi^2 J},$$

and thus to introduce the mysterious "half-quanta" that would plague quantum theory until they were justified by the new quantum mechanics in 1925-1926.[13] According to Reiche and those who followed him, rotationfree states of molecules did not exist.

From the perspective of the present paper another feature of Imes's data is more important. He found that each of the peaks in the 1.76µ band could be further divided in two peaks, with the weaker component being on the long wavelength side (Figure 2). Imes measured the doublet width to $\Delta\lambda = 14 \pm 1$ Å and remarked: "The apparent tendency of some of the maxima to resolve into doublets in the case of the HCl harmonic may be due to errors of observation, but it seems significant that the small secondary maxima are all on the long-wave side of the principal maxima they accompany" (Imes, 1919, p. 275).

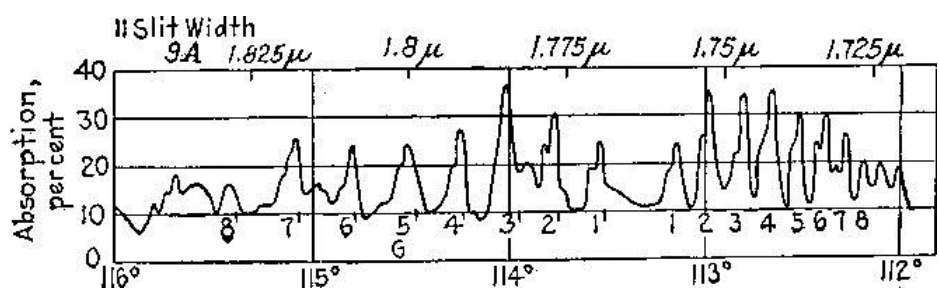

Figure 2. Imes's recording of the 1.76µ band of HCl.

Not knowing about isotopes of chlorine, Imes did not associate his observation to a possible isotope effect. However, it so happened that on 18 December 1919 Aston announced that he had succeeded in separating the element by means of his new mass spectrograph. According to Aston, chlorine was a mixture of two isotopes Cl-35 and Cl-37, possibly with a trace of a third Cl-39 isotope (Aston, 1919; 1920). Given that the atomic weight of chlorine was 35.46 it followed that the abundance of Cl-37 was about 25%. Harkins, who for some time had been occupied by separating chlorine by means of a diffusion method, claimed to have found the isotopes earlier than Aston (Harkins, 1920). Constantly involved in priority disputes, it was not without reason that he was nicknamed "priority Harkins."

---

[13] Reiche (1920) attributed the idea of half-quanta to Einstein. On half-quanta and zero-point energy in the old quantum theory, see Mehra and Rechenberg (1999) and Gearhart (2010).

The discovery of the isotope effect in 1920 was what sociologist of science Robert Merton called a "doublet," a discovery made independently and nearly simultaneously by two scientists or research groups (Merton, 1973, pp. 343-370). Aware of the curious humps in Imes's absorption spectra of HCl and also of Aston's resolution of chlorine into isotopes, Francis Wheeler Loomis at New York University and Adolf Kratzer at the University of Göttingen both realized that Imes's data could be explained in terms of an isotope shift. Loomis, who was slightly ahead of Kratzer, announced his conclusion in the issue of *Nature* of 7 October 1920 while Kratzer's paper in *Zeitschrift für Physik* was dated 28 November (Loomis, 1920a; 1920b; Kratzer, 1920). Originally unaware of Loomis's note, in a postscript of 3 March 1921 to his earlier work Kratzer (1921) acknowledged the priority of his American colleague and rival. For the sake of completeness it should be mentioned that also the Austrian physicist Arthur Erich Haas (1921, dated 20 November 1920) came to the conclusion that the chlorine isotopes might turn up in molecular spectra in the infrared region. However, Haas was unaware of Imes's spectra and thus could provide no experimental support for his hypothesis. The discovery was a doublet, not a triplet.

The approaches followed by Loomis and Kratzer were essentially the same and can be summarized as follows. Considering only the vibrational term, they both assumed that the force constant $k$ appearing in the frequency of a vibrating electron,

$$\nu = \frac{1}{2\pi}\sqrt{\frac{k}{\mu}}$$

would not be affected by an isotopic substitution. For the reduced mass μ in terms of the hydrogen mass Loomis wrote it as µ′ = 35/36 and µ″ = 37/38, referring to the light and heavy chlorine isotope, respectively. Then

$$\frac{\mu''}{\mu'} = 1 + \frac{2}{1330},$$

from which followed with good approximation

$$\frac{\nu' - \nu''}{\nu'} = 1 - \sqrt{\frac{\mu'}{\mu''}} = \frac{1}{1330}$$

For HBr he obtained a similar result, only with 6478 as denominator. He concluded that "the absorption spectrum of ordinary HCl should consist of pairs of lines separated by 1/1330 of their frequency" (Loomis, 1920b, p. 253). This corresponded to a difference in wavelength of 13 Å, in satisfactory agreement with the measurements reported by Imes.

Kratzer's analysis was more sophisticated than Loomis's, but the general idea was the same. For the relative difference between the vibrational frequencies in HCl³⁵ and HCl³⁷ he derived

$$\frac{\Delta\nu}{\nu} = -\frac{1}{2}\Delta\left(\frac{1}{\mu}\right) = \frac{1}{2}\left(\frac{1}{\mu'} - \frac{1}{\mu''}\right) = \frac{1}{1295}$$

At $\lambda = 1.76\mu = 1.76 \times 10^{-4}$ cm, this expression transforms into the difference in wavelength:

$$\Delta\lambda = -\frac{\Delta\nu}{\nu}\lambda = \frac{1.76}{1295} \times 10^{-4} \text{ cm} = -13.54 \text{ Å},$$

where the minus sign indicates that the wavelength of the Cl-35 component is smaller than the one of the Cl-37 component. Moreover, the ratio of measured intensities corresponded roughly to the expected abundance ratio. The agreement with Imes's experiment was perfect. In the opinion of Sommerfeld (1922, p. 520), the papers by Loomis and Kratzer provided "the most beautiful confirmation of Aston's views of isotopy."

Some years later Loomis wrote a comprehensive review of the spectroscopic isotope effect in which he reconsidered his and Kratzer's work as well as other developments in the field (Loomis et al., 1926, pp. 260-271). By that time Robert Mulliken at Harvard University had carried out the first detailed study of isotope effects in the electronic bands of diatomic molecules. In an investigation of the spectrum of boron monoxide, involving two systems due to $B^{10}O$ and $B^{11}O$, he argued convincingly for the reality of the previously discussed zero-point energy (Mulliken, 1925; Mehra and Rechenberg, 1999).

## 6. The discovery of deuterium

Whereas the isotope effect was established for molecules by 1920, it took more than another decade until the corresponding atomic effect was confirmed. This happened in connection with the Nobel Prize-rewarded discovery of deuterium, the heavy hydrogen isotope.[14] Two German physical chemists, Otto Stern and Max Volmer (1919), had searched for the heavy isotope by means of diffusion methods, but were led to the disappointing conclusion that it was less abundant than in the ratio 1 : 100 000. Nor did other searches succeed. In his monograph *Isotopes*, Aston (1923, p. 70) stated confidently that "hydrogen is a simple element," a claim there seemed no reason to doubt.

Only in 1931 did Raymond Birge and Donald Menzel refer to the possibility of a heavier hydrogen isotope, which they thought might explain why the atomic weight of the element was the same on the physical and chemical mass scale.[15] The coincidence might be explained, they suggested, "by postulating the existence of an isotope of hydrogen of mass 2, with a relative abundance $H^1/H^2$ = 4500" (Birge and Menzel, 1931, p. 1670). They thought it should be "possible, although difficult, to detect such an isotope by means of band spectra." Inspired by the idea of Birge and Menzel, Harold Clayton Urey at Columbia University decided to look systematically for a heavy isotope of hydrogen. The search, which he undertook in collaboration with his colleague George Murphy at Columbia, focused on the atomic rather than the molecular spectrum. For supply of liquid hydrogen rich in the supposed heavy isotope they teamed up with Ferdinand Brickwedde at the

---

[14] For the discovery of deuterium, see Murphy, 1964, and Brickwedde, 1982, both written by participants in the discovery. See also Dahl, 1999, pp. 22-27. There is no scholarly study of the discovery.

[15] The physical mass scale was based on $O^{16}$ = 16, while according to the chemical scale naturally occurring oxygen was assigned atomic weight 16. In 1935 it turned out that the "prediction" of Birge and Menzel was based on an incorrect atomic weight of hydrogen. With the new "physical" weight of 1.0081 instead of 1.0078, the argument of Birge and Menzel lost its validity, such as Urey pointed out in a postscript to his Nobel Lecture (Brickwedde, 1982).

National Bureau of Standards in Washington D.C. The samples provided by Brickwedde were prepared by means of fractional distillation.

| Spectral line | $\lambda_H$ | $\lambda_D$ | $\lambda_H - \lambda_D$ calculated | $\lambda_H - \lambda_D$ observed |
|---|---|---|---|---|
| $\alpha$ (3 → 2) | 6564.686 | 6562.899 | 1.787 | 1.79 |
| $\beta$ (4 → 2) | 4862.730 | 4861.407 | 1.323 | 1.33 |
| $\gamma$ (5 → 2) | 4341.723 | 4340.541 | 1.182 | 1.19 |
| $\delta$ (6 → 2) | 4102.929 | 4101.812 | 1.117 | 1.11 |

Table 1. Wavelengths in Å for Balmer lines. Columns 2-4 are theoretical values according to Bohr's theory, while column 5 gives the mean differences in wavelength observed by Urey and Murphy.

Even with ordinary hydrogen Urey and Murphy detected faint lines on their plates that corresponded to the calculated positions of the β, γ, and δ lines in the Balmer spectrum of the mass-2 isotope (corresponding to transitions from stationary states $n$ = 4, 5, and 6 to $n$ = 2). Using the samples with increased concentration of deuterium they found that the intensity of the Balmer lines increased. Importantly, Urey and Murphy could now measure the $D_\alpha$ line and observe its doublet structure, the width of which turned out to be of the same order as the $H_\alpha$ fine structure doublet. While the latter was known to be $\Delta\lambda$ = 0.135 Å, for deuterium they found $\Delta\lambda \cong 0.11$ Å. As to the relative abundance they estimated D : H = 1 : 4000, in good agreement with the Birge-Menzel estimate and of the same order as the presently known ratio 1 : 6700. The two Columbia physicists also searched for the H-3 (tritium) isotope, but found none. Although "no evidence for $H^3$ has yet been found, … [it] may yet show that this nuclear species exists" (Urey et al., 1932c, p. 15; see also Lewis and Spedding, 1933).

Urey, Murphy, and Brickwedde first reported their results at the annual meeting of the American Physical Society at Tulane University, New Orleans, on the last days of 1931. More details followed in a note of January 1932, and the full report appeared in the *Physical Review* in its issue of 1 April 1932 (Urey et al., 1932a; 1932b; 1932c). The discovery was quickly confirmed by Walker Bleakney (1932) at Princeton University, who in a mass-spectrographic study detected the molecular ion $HD_2^+$ as distinct from the $H_3^+$ ion. Two years later Urey was awarded the Nobel Prize in chemistry "for his discovery of heavy hydrogen." Remarkably, in none of their papers of 1932 did Urey and his coauthors refer to their work on deuterium as a "discovery."

## 7. Discussion

The Bohr atomic model and the concept of isotopy both date from 1913. Although the former was primarily a theory of the electronic part of the atom, and the latter is a nuclear property, from a historical point of view the two conceptions were related. Bohr was the first to realize, and in this sense to predict, that isotopes of the same element will exhibit slightly different spectra. However, his prediction was not generally known in the physics community.

While the atomic isotope effect was only confirmed in 1932, when deuterium was discovered spectroscopically, the molecular effect was established as early as 1920. The effect discovered independently by Loomis and Kratzer was of a different kind than the one originally conceived by Bohr, as it did not involve quantum transitions between stationary states. The relationship between theory and experiment is interesting: Loomis and Kratzer did not really *predict* the effect in HCl since the data in the form of Imes's spectra already existed; relying on Aston's resolution of chlorine in two isotopes, they *interpreted* the data as an isotope effect and thus confirmed that a nuclear property can affect the spectra of molecules. In the case of Urey's discovery of deuterium, the theory came many years before the experiment and directly guided the Columbia research programme.